\documentclass[%
superscriptaddress,
 amsmath,amssymb,
 aps,
 reprint,
prb,
floatfix,
]{revtex4-2}
\usepackage{graphicx,xcolor}
\usepackage{multirow}
\definecolor{darkblue}{RGB}{0,0,150}
\definecolor{nightblue}{RGB}{0,0,100}
\newcommand{\matpro}{\textsc{Materials Project} }

\usepackage{mathrsfs,dsfont,mathtools}

\usepackage{ulem}

\usepackage{dcolumn}
\usepackage{bm}
\usepackage[
colorlinks,
citecolor=darkblue,
linkcolor=darkblue,
urlcolor=nightblue]{hyperref}

\usepackage[english]{babel}
\usepackage[babel,kerning=true,spacing=true]{microtype}
\usepackage{pifont}
\usepackage{feynmp-auto}

\makeatletter
\newcommand{\btriangle}{\mathpalette\btriangle@\relax}
\newcommand{\btriangle@}[2]{%
  \begingroup
  \sbox\z@{$\m@th#1\triangle$}%
  \makebox[\wd\z@]{%
    \raisebox{0.04\height}{%
      \resizebox{1.1\wd\z@}{0.96\ht\z@}{%
        $\m@th#1\blacktriangle$%
      }%
    }%
  }%
  \endgroup
}
\makeatother

\AtBeginDocument{\renewcommand{\natexlab}[1]{#1}}
\begin{document}

\title{
Spatiotemporal Order and Parametric Instabilities from First-Principles
}
\author{Daniel Kaplan}
\email{d.kaplan1@rutgers.edu}
\affiliation{Center for Materials Theory, Department of Physics and Astronomy, 
Rutgers University, Piscataway, NJ 08854, USA}%
\author{Pavel A. Volkov}
\affiliation{Department of Physics, University of Connecticut, Storrs, Connecticut 06269, USA}
\author{Jennifer Coulter}
\affiliation{Center for Computational Quantum Physics, The Flatiron Institute,162 5th Avenue, New York, NY 10010}
\author{Shiwei Zhang}
\affiliation{Center for Computational Quantum Physics, The Flatiron Institute,162 5th Avenue, New York, NY 10010}
\author{Premala Chandra}
\affiliation{Center for Materials Theory, Department of Physics and Astronomy, 
Rutgers University, Piscataway, NJ 08854, USA}
\begin{abstract}
Shaping crystal structure with light is an enduring goal of physics and materials engineering.
Here we present calculations in candidate materials selected by symmetry that allow light-induced spatiotemporal parametric instabilities. 
We demonstrate a theoretical framework that includes a complete symmetry analysis of phonon modes that contribute to parametric instabilities across all non-centrosymmetric point groups, a detailed survey of the materials landscape and finally the computation of nonlinear couplings from first principles. We then showcase detailed results for chiral crystals, ferroelectrics, and layered van der Waals materials. 
Our results pave the way towards realizing designer time-crystalline order in quantum materials, detectable with time-resolved diffractive probes.
\end{abstract}

\maketitle

\section{Introduction}
The possibility of designing light-induced periodic structures in quantum materials has motivated theoretical and experimental research \cite{disa2021engineering}. If realized, such exotic light-matter interaction could lead to switchable, reversible control of symmetry-breaking  structural transitions \cite{Kaplan2025}. Such ``designer" structures on nanometer scales might result in exotic electronic phases and would have applications in memory storage, computational devices and optoelectronics.
\begin{figure}[!ht]
    \centering
    \includegraphics[width=1.0\columnwidth]{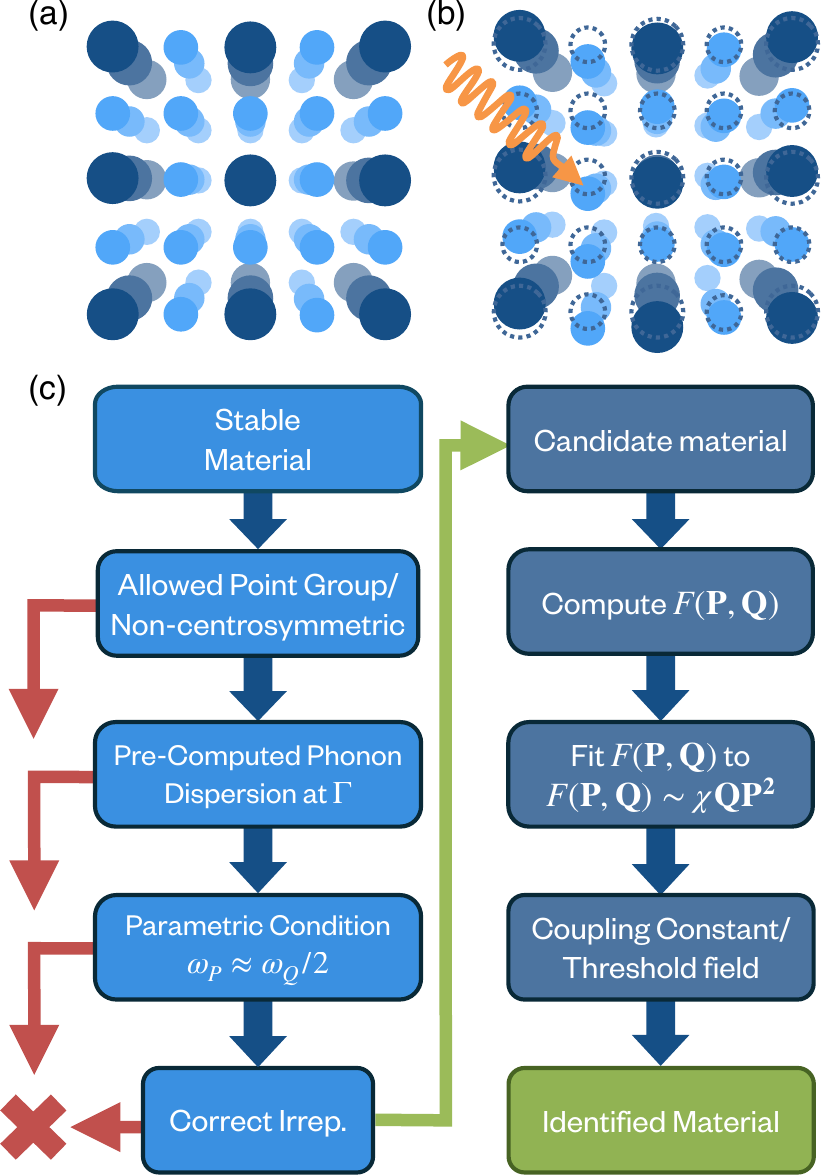}
    \caption{Schematic illustration of our theoretical framework and workflow for materials search. Panel (a) shows the equilibrium crystal, and (b) its light-driven state. In (c) the workflow is displayed. The left (light blue boxes) are screening steps from database, followed by computation steps on the right (dark boxes). A successful candidate is marked with green.}
    \label{fig:intro}
\end{figure}
A key challenge in developing photoinduced incommensurate structures is that light transfers negligible momentum to massive particles, typically leading to instabilities only on long wavelength scales.

Traditionally, this has meant that finite-momentum modes in materials have been accessible through mechanisms such as (a) Raman scattering or (b) incoherent down-conversion of high-energy pulses through higher-order nonlinearities in crystals . Both of these approaches have disadvantages: (a) is typically significantly weaker than the standard dipolar light-matter coupling and thus is inefficient; 
(b) leads to the generation of an ensemble of momenta, without clear signatures of order.  Furthermore this incoherent decay process excites many finite-momentum modes of different symmetries, and thus would not result in tunable forms of spatial ordering. 

Here we highlight the existence of another mechanism, {\it parametric resonance}, an enhanced form of down-conversion promoted by nonlinearities at specific wavevectors \cite{landau2013mechanics}. This phenomenon manifests itself in a \textit{coherent} energy transfer from an excited mode at frequency $\Omega$ to a mode pair at momenta and frequency $(\pm q_0,\Omega/2)$. Parametric resonance has a long history in physics \cite{landau2013mechanics} and has been explored in many contexts \cite{Yao20,Zaletel23} including
surface waves \cite{Agam01}, magnonics \cite{Demokritov06, hosseinabadi2025,kiselev2025}, plasmonics \cite{Kiselev2024}, Josephson junctions \cite{Kleiner21} and
Cooper pair turbulence in atomic gases \cite{Dzero09}. Parametric resonance in solids has been primarily studied  through incoherent excitations,
giving light access to finite-momentum ``silent" modes \cite{Orbach66,Liu13,Cartella18} and through direct coupling \cite{Juraschek20}. 

In a recent paper (Ref. \cite{Kaplan2025}), some of us have demonstrated that light-driven phonons with nonlinear interactions exhibit parametric responses, leading to states with broken space- and time- translational symmetry as  
schematically presented in Figs.~\ref{fig:intro} (a)-(b); these spatial and temporal patterns are controlled by the drive frequency $\Omega$ and the dispersion of the irradiated material.
This emergence of coherent finite-momentum spatiotemporal ordering, tunable by light fluence, has been identified using a combination of analytic and numerical approaches. However, the precise symmetry restrictions on this effect in three dimensions, as well as examples of real crystals that could host it, were not clarified. 

Here we perform a focused search for promising materials to host such spatiotemporal order via parametric resonance. In the process, we answer a number of unresolved questions related to the experimental identification of these materials, providing symmetry restrictions on the proposed parametric coupling, predictions of its magnitude, and strategies to enhance it.  We address these issues by conducting a comprehensive search for candidates with symmetry-allowed light-induced structural instabilities using open-source data provided by the \textsc{Materials Project} \cite{jain2013commentary,munro2020improved,petretto2018high}.
In Fig.~\ref{fig:intro}(c) we present the workflow we
established to identify materials with strong nonlinear couplings within specific constraints: we begin by considering materials that break inversion symmetry and 
 thus have chiral or polar point group symmetries; subsequently we narrow the search further to include constraints on available phonon energies as well as other important conditions required to realize parametric instabilities.

The structure of this paper is as follows: in Sec.~\ref{sec:theory} we present the theory underlying our search. Next we describe in Sec.~\ref{sec:mat_search} the methodology involved, indicating
the specific work flow in our materials pipeline.  We present results in Sec.~\ref{sec:res}; this is followed by Sec.~\ref{sec:disc} where we summarize
our findings 

and highlight a set of materials which may exhibit light-driven spatiotemporal order,
mention experimental fingerprints and
suggest directions for future work.

Our work serves as a critical first step in the realization of spatiotemporal order shaped by light in real materials, which will permit hosting out-of-equilibrium tunable incommensurate potentials with arbitrary periodicity.

\section{Theory}
\label{sec:theory}
The parametric instability discussed here consists of a light-coupled phonon $\mathbf{Q}(\mathbf{r},t)$ and a parametrically driven mode $P(\mathbf{r},t)$. A general, local coupling between these modes is given by,
\begin{align}
V = \sum_{n=1}^{\infty} \chi^{(n_1,n_2)}_{i_1 i_2  \ldots i_{n_1} j_1 j_2\ldots j_{n_2}} Q_{i_1}Q_{i_2} \ldots P_{j_1}P_{j_2}\ldots,
\label{eq:general_coupling}
\end{align}
where $(n)$ is the order of the coupling and $i_1 \ldots j_1$ are Cartesian indices of the phonon modes. 
We assume the material in question is stable in equilibrium and we therefore treat the series in Eq.~\eqref{eq:general_coupling} to be perturbative and truncatable at a finite order. We note that including additional terms, in the perturbative sense, does not qualitatively affect the possibility of an instability.
The fundamental restriction thus far placed on the modes concerns the property of $\mathbf{Q}(\mathbf{r},t)$ being resonantly light-driven. This means $\mathbf{Q}$ \textit{must} transform as a dipole in any point group and specifically as a vector \textbf{odd} under inversion, when inversion is part of the symmetry group. In addition, for coupling to a transverse field such as light with associated electric field $\mathbf{E}$, we require that $\mathbf{Q}$ be an optical transverse mode. The equations of motion for $\mathbf{Q}(\mathbf{r},t), \mathbf{P}(\mathbf{r},t)$ read,
\begin{align}
    \partial_t^2 Q_i-v^2 \nabla^2Q_i +\omega_Q^2Q_i + \beta\partial_tQ_i + \frac{\delta V}{\delta Q_i}  &= \frac{Z_{i}^{Q} E_i(t)}{M_Q},\label{eq:force} \\
     \partial_t^2 P_i-v^2 \nabla^2P_i +\omega_P^2P_i + \beta\partial_tP_i + \frac{\delta V}{\delta P_i} &=0
\end{align}
Here $v^2$ is the velocity of the respective modes, $\omega_{P/Q}^2$ is the mass of the mode, $\beta$ is the dissipation, $Z^{Q}$ is the effective charge and $V$ is the local coupling between modes expressed in Eq.~\eqref{eq:general_coupling}. 
For simplicity, consider a 1D model with phonon dispersion $\omega^2(q) \approx \omega_0^2 + v^2 q^2$. The results are straightforward to generalize for arbitrary $\omega(q)$.

This equation is conveniently analyzed in momentum and frequency space. Performing the Fourier transform, we  evaluate $Q_i(\mathbf{q},\omega)$. Assuming a harmonic drive, $E_i(t) = E^{0}_i\cos(\Omega t)$, $Q_i$ is resonant only for $\mathbf{q}=0$ and $\Omega \approx \omega_Q$. Thus, we shall assume the form of $Q_i \sim e^{\pm i\Omega t}$ and strictly at zero momentum $\mathbf{q}=0$. This imposes stringent contraints on the resultant coupling in $V$. Consider the lowest order term ($n_1=1,n_2=1$), expressed as its Fourier harmonics, $V_1 = \chi^{1}_{i_1j_1} Q_i(q_1,\omega_1)P_j(q_2,\omega_2) \delta(\omega_1+\omega_2)\delta(q_1+q_2)$. Since $q_1 = 0$ it follows that $q_2 = 0$ as well and no structural instability results from this term. The next order $n_1=1,n_2=2$ is the first non-trivial contribution. Here,
\begin{align}
    V_{1,2} = \chi_{i_1j_2j_3}^{1,2} Q_{i_1}(q_1,\omega_1)P_{j_2}(q_2,\omega_2)P_{j_2}(q_3,\omega_3).
\end{align}
Since $q_1 =0, \omega_1=\Omega$, we have $q = q_2=-q_3=-q$ and $\omega_2 + \omega_3 = \Omega$. Using time-reversal symmetry $\omega_2 = \omega_3 =\omega$. With this, we find a parametric resonance for a mode at frequency $\omega=\Omega/2$ and momentum $q$, which is fixed by the dispersion relation for $\mathbf{P}(\mathbf{r},t)$, $\omega_P(q) = \Omega/2$. Following the above reasoning we discard the term of equal $n_1+n_2$ order, i.e., $n_1=2,n_2=1$. This contribution is of the form $Q_{i_1} Q_{i_2}P_{j_1}$. Since $Q$ is purely at $q=0$, $P$ is thus constrained to be at $q=0$ as well, precluding the possibility of any instabilities. A term of the form $P^2 Q^2$ is discarded for reasons discussed in Sec.~\ref{sec:disc}.

Other higher order contributions are smaller by construction and therefore omitted here.
Importantly, this cubic form allows for an analytically tractable solution for the strength $E_i^{0}$ that leads to an instability \cite{Kaplan2025}; labeling this as $E_c$ (critical field), we find,
\begin{align}
    \frac{Z ^{Q} E_c}{M_Q M_P} = \frac{(\beta \Omega )^2}{2|\chi^{1,2}|}.
    \label{eq:coupl_define}
\end{align}

We mention here that beyond the parametric coupling $V_{1,2}$, higher order couplings e.g., of the form $V_{1,3} = Q_{i_1}P_{j_1}P_{j_2}P_{j_3}$, may be allowed in principle, and lead to down-conversion. From the stand point of perturbation theory, they are expected to be smaller than $V_{1,2}$. In order to satisfy the energy and momentum conservation constraints, $\sum_{i=1}^3 \mathbf{q}_i =0$ and $\sum_{i=1}^3 \omega_i = \Omega$ (where $\mathbf{q}_i , \omega_i$ are the wavevector and frequency of the $P_{j_i}$ mode), nonlinearities and other non-circular properties of the phonon dispersion will play a key role in determining whether these conditions are met. This is in contrast to the $V_{1,2}$ studied here, where the situation is more transparent. Only pairs of modes with wavevectors $\mathbf{q}$ and -$\mathbf{q}$ are parametrically excited in that case. As $\omega_P({\bf q}) = \omega_P(-{\bf q})$ due to time-reversal symmetry, the parametric resonance condition reduces to $\omega_P({\bf q}) =\Omega/2$. We thus focus on the effects of $V_{1,2}$ in this work and leave investigation of higher-order coupling to future work.
Beyond momentum and energy conservation, crystal symmetry restricts the components of the coupling tensor $\chi^{1,2}_{i_1j_2j_3}$. We first divide all point groups to those containing the inversion and those without it. For groups containing inversion, $Q_{i_1}$ must transform \textit{ungerade}. However, for $P$ for any symmetry $P_{j_1}P_{j_2}$ always transforms \textit{gerade}. Since the product of all three $Q P P$ is odd under inversion, this coupling vanishes identically. Thus, inversion must be broken for a parametric coupling to exist.

Out of the 32 crystallographic point groups, 21 lack a center of inversion \cite{butler2012point}. We focus on the latter, in which driven parametric instabilities are expected to arise. The remaining point group restriction stem from the need of the triple product to lie within the trivial representation $A$. For each of the sixteen non-centrosymmetric point groups, we enumerate the irreps. of the modes which may couple leading to a parametric down-conversion.
\begin{table*}[!ht]
\begin{tabular}{ |p{3cm}|p{3cm}|p{3cm}|p{4cm}| p{2.8cm} |}
 \hline
 \multicolumn{5}{|c|}{Point groups and mode symmetries} \\
 \hline
  Crystal class & Point group & $\mathbf{Q}$ irrep. & $\mathbf{P}$ irrep. & Example\\
 \hline
 \multirow{1}{*}{Triclinic}    & 1     &All &   All & KIO\textsubscript{3} \\ \hline
 \multirow{2}{*} {Monoclinic} &   2   & \textit{A}   & All  & HgNO\textsubscript{3} \\ \cline{2-5}
  & m& \textit{A}$'$&  All  & Al$_2$Se$_3$\\ \cline{1-5} 
  \multirow{2}{*} {Orthorhombic}    & mm2  & \textit{A}$_1$&  $
  A_1 \oplus B_1 \oplus B_2$ & CoAsS (CrSBr) \\ \cline{2-5}
  & 222  & $B_1 \oplus B_2 \oplus B_3$ &  \ding{55} & \ding{55} \\ \cline{1-5}
    \multirow{3}{*}{Tetragonal}  & 4 & $A$ & All  & Cs\textsubscript{3}P$_7$\\ \cline{2-5} 
    & $-4$ & $B$ & $E$ & InPS$_4$ \\ \cline{2-5} 
    & 4mm & $A_1$ & All & PbTiO$_3$ \\ \cline{2-5} 
    & -42m & $B_2$ & $E $ & NaPN2 \\ \cline{2-5} 
    & 422 & $A_2\oplus E$ & \ding{55} & \ding{55} \\ \cline{1-5} 
  \multirow{3}{*}{Trigonal} & \multirow{2}{*}{3} & $A$ & All & CsNO$_3$ \\ \cline{3-4} 
&  & $E$ & $E$ & \\ \cline{2-5}
& \multirow{2}{*}{3m} & $A_1$ & All & BaTiO$_3$ (T=0) \\ \cline{3-4} 
&  & $E$ & $E$ & \\ \cline{2-5}
& 32 & $E$ & E & Se \\ \cline{1-5}
\multirow{5}{*}{Hexagonal} & 6  & A & All & BaAl$_2$O$_4$ \\ \cline{2-5} 
& -6 & $E'$ & $E' \oplus E''$ & NaLiCO$_3$ \\ \cline{2-5}
& 6mm & $A_1$ & All & ZnS (wurtzite) \\ \cline{2-5}
& -62m  & $E'$ & $E' \oplus E''$ & CaP \\ \cline{2-5}
& 622 & $A_2\oplus E_1$ & \ding{55} & \ding{55} \\ \cline{1-5}
 \multirow{3}{*}{Cubic} & 23  & T &T & RuSi\\ \cline{2-5}
 & 432  & $T_1$  & \ding{55} & \ding{55}\\ \cline{2-5}
  & -43m  & $T_2$  & $T_2$& LiZnP (zincblende)\\ 
 \hline
\end{tabular}

 \caption{ \label{tab:1} Point group symmetries and the parametric coupling. For every one of the 21 non-centrosymmetric point groups, we enumerate the representation of the IR mode $\mathbf{Q}$ that drives a parametric instability. We list only irreps. of $\mathbf{Q}$ that may couple to the square of the mode $\mathbf{P}$; the irreps. of $\mathbf{P}$ that lead to an instability are listed. A material example for each point group is provided. In some entries, we list the dipole irreps. of $\mathbf{Q}$ but indicate that there is no allowed down-converted coupling by \ding{55}.  }
\end{table*}

Tab.~\ref{tab:1} exhaustively lists all crystallographic point groups and the requisite irreps. of the IR mode $\mathbf{Q}$ and the down-converted mode $\mathbf{P}$ for all noncentrosymmetric point groups. Several features are immediately apparent: for the lowest symmetry groups $1,2,\textrm{m}$, virtually all phonons may participate in down-conversion. Naturally, higher symmetry groups result in a lower propensity to instability consistent with naive expectations \cite{dunitz1996symmetry}.

We note that genuine ferroelectricity or a net polar moment is not required for parametric coupling; we find several point groups without a center of inversion -- that are otherwise non-polar -- which permit such instabilities. Examples include elemental compounds, such as Se and Te. 
In addition to the restriction of point group symmetries, we note that the parametric instability requires that the drive IR mode frequency $\omega_Q=\Omega = 2\omega_P(q)$ is twice the frequency of the down-converted mode at momentum $q$. Note that in 3D systems there can be, in general, a surface in ${\bf q}$ space, satisfying this condition; however, as nonlinearity and damping have momentum dependence determined by the discrete point group of the material, we expect only a finite number of possible instabilities.

If there are no modes satisfying the exact parametric condition of $\omega(\mathbf{q})=\Omega/2$ can be found, it has been shown in Ref.~\cite{Kaplan2025} that this constraint may be somewhat relaxed at the cost of a higher onset field for the parametric instability.

In the next section, we review the 
focused search conducted through known materials towards selecting candidates for parametric instabilities. 

\section{Materials Search}
\label{sec:mat_search}
At present, there exist databases with a wide variety of deposited crystal, electronic and vibrational properties of solids. To navigate through the materials landscape, we choose \textsc{Materials Project} \cite{jain2013commentary}, for its compatibility with modern first-principles codes and APIs (Application Program Interfaces) that permit constrained and efficient scans of the database. 
\begin{figure}[!h]
    \centering
    \includegraphics[width=0.8\columnwidth]{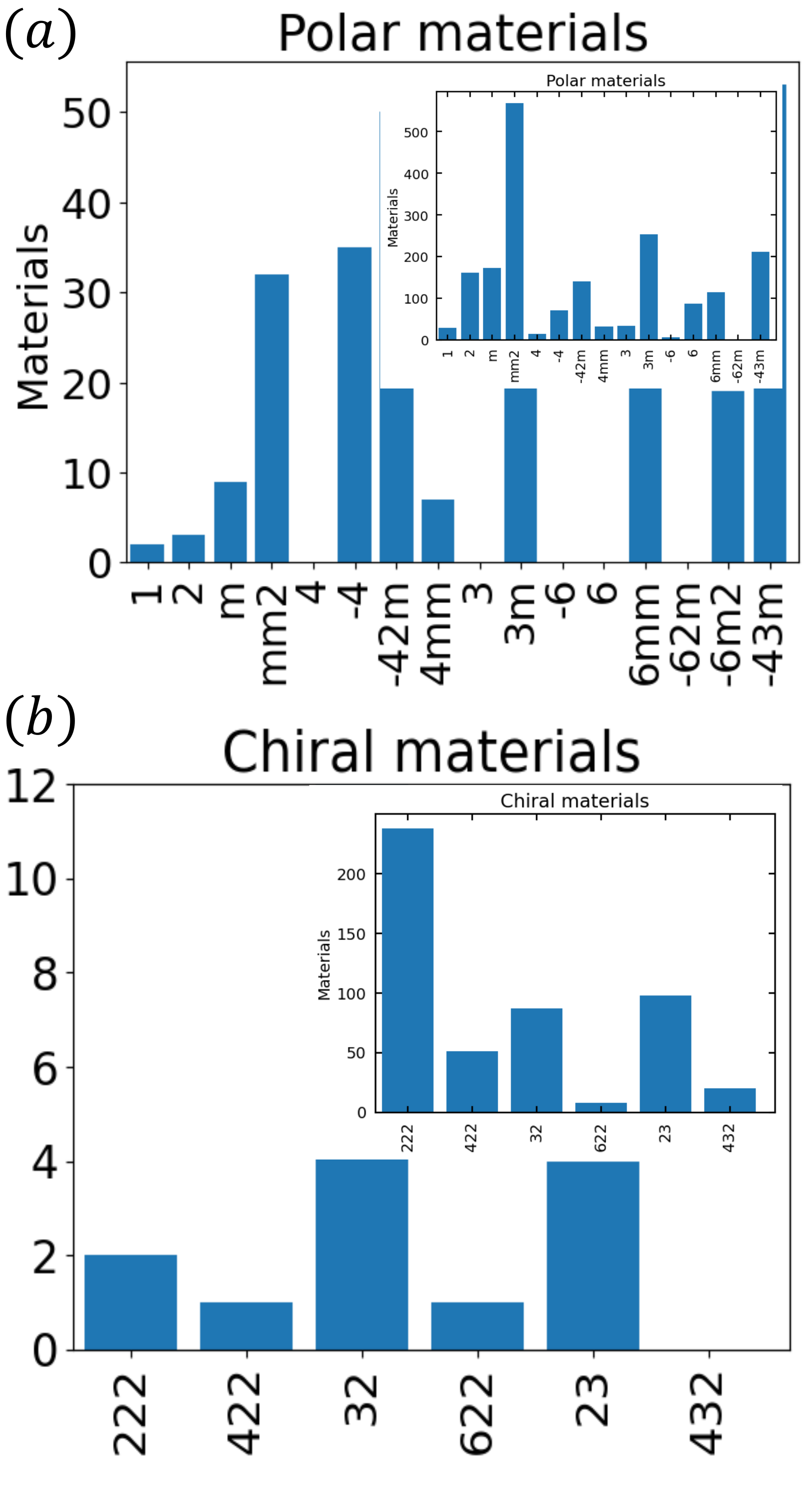}
    \caption{Distribution of materials with parametric instabilities with (a) polar and (b) chiral point symmetries; insets show the full number of materials found and the larger
    histograms represent the subset of those materials with existing phonon bandstructures.}
    \label{fig:survey_1}
\end{figure}
To make sure our results remain experimentally relevant, we focus exclusively on stable materials. This corresponds to narrowing down the pool of candidates to those with an energy above the convex hull $E_h < 10^{-3} ~\textrm{eV}$. From those, we pick out the ones belonging to \textit{all} non-centrosymmetric point groups as listed in Tab.~\ref{tab:1}. We restrict the search to non-magnetic insulators, specifically with the intent of allowing light to couple only to low energy degrees of freedom such as phonons. From the 35984 stable materials currently featured in 
\textsc{Materials Project}, only 2078 fit the above mentioned criteria and are within the 21 point groups which may in principle allow parametric instabilities. 

The statistics of this collection is presented in the insets of Figs.~\ref{fig:survey_1}(a)-(b). Here, we separate the classes of materials for parametric instability into two categories: polar and chiral (i.e., non-centrosymmetric but non-polar). In order to leverage existing, high-quality phonon calculations on the full database of materials, we restrict our search to those systems which have a phonon dispersion pre-computed and deposited in \cite{petretto2018high}. This additional filtration results in 278 candidates, whose distribution is presented in Fig.~\ref{fig:survey_1}(a)-(b) (main panel). We find a roughly even blend of materials across point groups, with notable contributions coming from orthorhombic (mm2), trigonal (3m) and cubic ($\bar{4}$3m) point groups; this may be due to their relative structural simplicity and higher symmetry. In the overall database of \matpro (insets), we find relatively similar distribution of point groups as in the smaller set we have scanned, suggesting that we collected a representative sample. 

The process for identifying a viable candidate is as follows:
\begin{enumerate}
    \item The material stable is stable, $E_h < 10^{-3} \textrm{eV}$;
    \item The material is non-centrosymmetric, and is gapped (within PBE/GGA, as reported in Ref.~\cite{jain2013commentary});
    \item The material belongs to one of the 21 point groups relevant for parametric instabilities;
    \item The material has pre-computed phonon dispersion (with dynamical matrices) deposited in Ref.~\cite{petretto2018high};
    \item Label the phonon mode symmetries at $\Gamma$;
    \item In keeping with Tab.~\ref{tab:1}, check whether there are two modes with irreps. that conform to an IR mode and down-converted mode with frequencies s.t. $\omega_{Q} = 2\omega_P \pm 0.2\omega_p$;
\end{enumerate}

After identifying the material in this manner, we proceed to compute the coupling constant. The energy of the system in equilibrium is determined by its Born-Oppenheimer surface \cite{woolley1976quantum,essen1977physics,argaman2000density}. The total energy of the system,
\begin{align}
    E = E(\lbrace\mathbf{R}_i\rbrace),
\end{align}
where $\mathbf{R}_i$ are the atomic positions. In equilibrium, we require that $\frac{\partial E}{\partial R_{i\alpha}} = 0$, with $\alpha$ being the cartesian component of $\mathbf{R}_i$. Phonons are described by displacements vectors $u_{i,\alpha}$, $u_{i,\alpha} = R_{i,\alpha}-R_{i,\alpha}^{0}$. Here $\alpha$ is the cartesian component of the displacement of the $i$-th atom. 
The perturbation $u_{i,\alpha}$ is introduced for a self-consistent solution to linear order using DFPT (density functional perturbation theory) \cite{Gonze1997,Baroni2001}. The harmonic (``frozen") approximation yields \cite{weyrich1990frozen},
\begin{align}
    E = E(\lbrace \mathbf{R}^{0}_i\rbrace) + \sum_{i\alpha,j\beta}\frac{\partial^2 E}{\partial R_{i\alpha} R_{j\beta}} u_{i\alpha }u_{j\beta}. 
\end{align}
The dynamical matrix $D_{i\alpha j\beta} = \frac{\partial^2 E}{\partial R_{i\alpha} R_{j\beta}}$ can be then diagonalized to yield the normal modes of the solid. These yield the $\mathbf{Q}$ and $\mathbf{P}$ referred to, above. After defining the desired $\mathbf{Q}$ and $\mathbf{P}$ modes based on their respective symmetries, we displace the atoms,
\begin{align}
    \mathbf{R}_i \to \mathbf{R}_i+\lambda_1\langle \mathbf{Q},\hat{e}_i\rangle+\lambda_2\langle \mathbf{P},\hat{e}_i\rangle,
\end{align}
such that $\langle ...\rangle$ denotes the projection of the mode on the eigenvector $\mathbf{\hat{e}}_i$ for the i-th atom. $\lambda_{1,2}$ denote the physical amplitudes of the displacements along the $\mathbf{Q}, \mathbf{P}$ directions. From this, we compute the total energy and the coupling constant, 

\begin{align}
    \notag &E = E(\mathbf{R}_i^{0}) + E(\mathbf{R}_i^{0},\lambda_1,\lambda_2), \\
    &\chi^{1,2} = \frac{\partial}{\partial \lambda_1}\frac{\partial^2 E}{\partial\lambda_2^2}.
    \label{eq:coupling}
\end{align}
We note that such an ab-initio driven investigation of nonlinear phononics has already been successfully employed \cite{Subedi2014,Subedi2015,subedi2021light} particularly in the context of light-matter interaction. In Sec.~\ref{sec:theory}, the coupling in principle connects the $\mathbf{Q}$ mode at $\mathbf{q} =0$ with the $\mathbf{P}$ mode at finite momentum, such that $\omega_P(\mathbf{q}) = \omega_Q/2$. As a first-principles calculation for an incommensurate cell for an arbitrary $\mathbf{q}$ is unfeasible, we use the value of the coupling at $\mathbf{q} =0$, provided the frequency of the modes does not deviate significantly from the parametric resonance condition of $2 \omega_P=\omega_Q$. We fix a heuristic bound, identifying this mode as viable via $|\omega_P-2\omega_Q| <0.2\omega_P$. 
Computational details are presented  in App.~\ref{App:A}.

\begin{figure}[!ht]
    \centering
    \includegraphics[width=1.0\columnwidth]{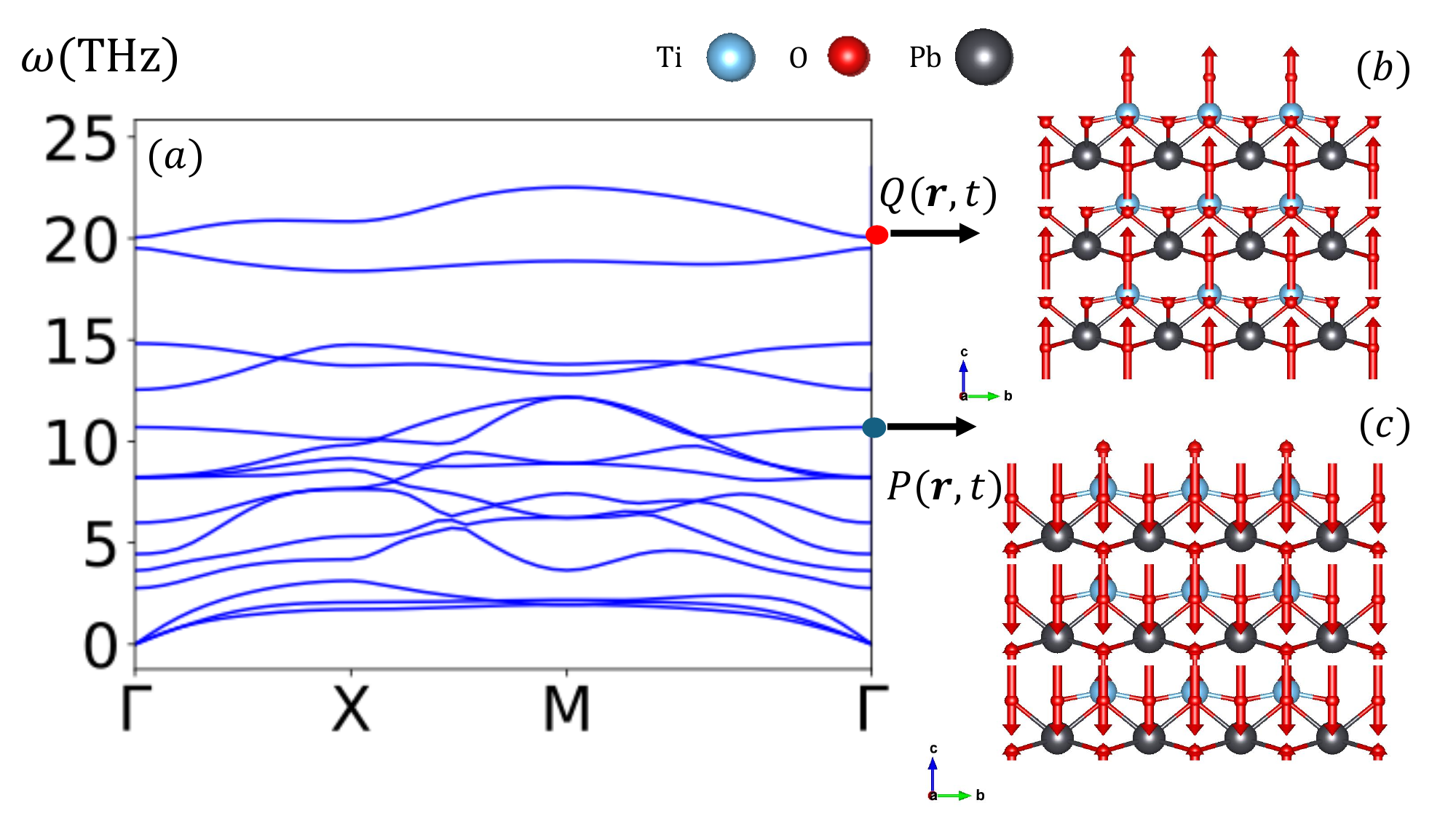}
    \caption{Phonon dispersion and parametrically down-converted phonon modes for PbTiO\textsubscript{3}. In (a), we show the phonon dispersion and identify the $Q$ mode with $A_1$ symmetry at $20.45 \textrm{THz}$. Within $\tfrac{\omega_Q}{2} \pm0.2 \omega_Q$, we find an $A_1$ mode at frequency $\omega_P = \textrm{11.052} \textrm{THz}$. According to Tab.~\ref{tab:1}, this coupling is permitted. The atomic displacements, for $Q$ and $P$ are shown in (b),(c), respectively.}
    \label{fig:fig2}
\end{figure}

 We present an overview of the methodology through the example of one of the candidates: tetragonal (4mm) PbTiO\textsubscript{3}. This polar ferroelectric \cite{schlom2007strain,rabe2007first,rabe2007modern,Ghosez1999,Tomeno2006,Freire1988} has been widely studied in the past, with the phonon dynamics in excellent agreement \cite{freire1981dynamical,Freire1988} with theory. Ref.~\cite{petretto2018high} computed the phonon dispersion for (4mm) PbTiO$_3$ which we plot in Fig.~\ref{fig:fig2}(a).
Using the procedure outlined above and Tab.~\ref{tab:1}, we identified the optically active mode ($A_1$ in this point group), as the phonon depicted in Fig.~\ref{fig:fig2}(b) at a frequency $\omega_Q = 20.45 \textrm{THz}$; in principle, this mode may couple to all others. The closest to the parametric condition $\omega_Q \approx 2 \omega_P$ was at $\omega_P = 11.04 \textrm{THz}$, and its symmetry was found to be $A_1$. 
\begin{figure}[!ht]
    \centering
    \includegraphics[width=1.0\columnwidth]{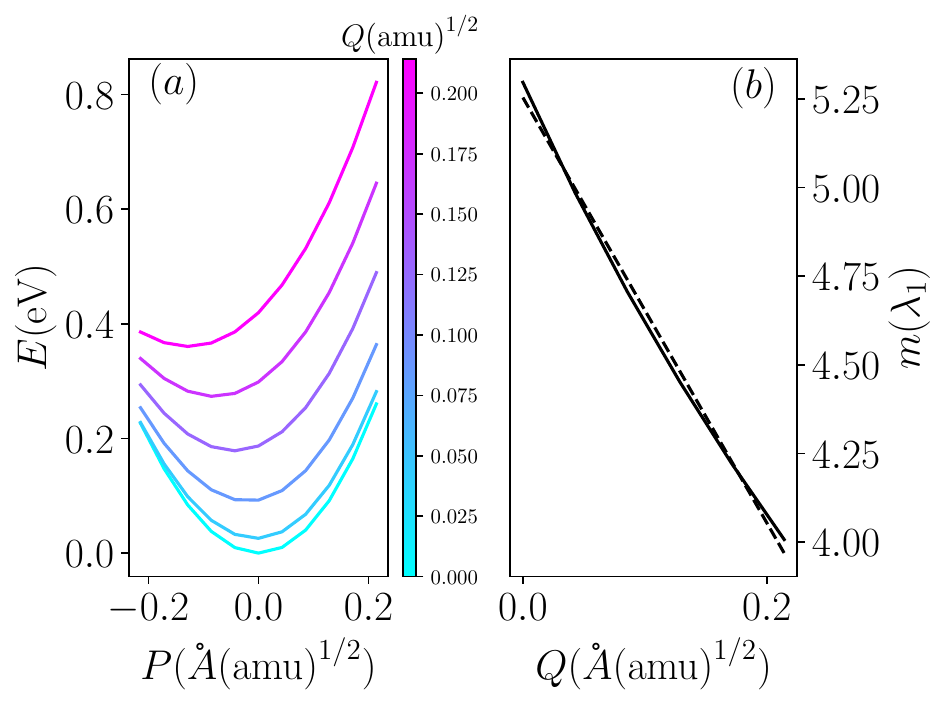}
    \caption{Calculation of the coupling constant. (a) Free energy $F$ as a function of mode displacements for every value of $\lambda_1$ ($Q$) and $\lambda_2$ ($P$). For simplicity, the free energy at $Q = 0, P=0$ is normalized to $0$. (b) Mode mass $\frac{\partial^2 F}{\partial \lambda_2^2}$ is given as a function of $Q$ amplitude ($\lambda_1$). Dashed line is a linear fit.}
    \label{fig:fig3}
\end{figure}
The calculation of the coupling constant proceeds as follows: We displace atoms in the conventional unit cell according to the normalized (to atomic mass) eigenmodes of $D_{i\alpha j\beta}$. The free energy $F$ is then re-evaluated for every displacement. Specifically, we calculate the mode mass, i.e., $\frac{\partial^2 F}{\partial \lambda_2^2}$ for every $|Q| \sim \lambda_1$. This is shown in Fig.~\ref{fig:fig3}(a). Thus, it follows that to leading order,
\begin{align}
    \frac{\partial^2 F}{\partial\lambda_2^2} = \omega^2_P+\lambda_1 m + \mathcal{O}(\lambda_1^2).
\end{align}
We fit the mode mass to a linear function, as shown in Fig.~\ref{fig:fig3}(b) (dashed line). Given the mode mass of primarily $O$ atoms ($\approx 16 \textrm{amu}$), we find a coupling constant $\chi^{1,2} = -25.87 \textrm{meV}/\AA^3$.
Separately, we have computed the Born effective charges for the $O$ atoms: $Z_O \approx -3.22$. 
For quantitative estimates, we shall throughout assume the broadening $\beta = 0.1\omega_Q \ll \omega_Q$.

Using the identity in Eq.~\eqref{eq:coupl_define}, we find the strength of the critical field $E_c = 4.68 ~MV \textrm{cm}^{-1}$ to be within current experimental capabilities in the THz regime \cite{basini2024terahertz,hafez2016intense,orenstein2025observation,fechner2024quenched}. From a broad survey of the phonon dispersion along high-symmetry lines, we determine the most likely ordering vector, that is the folding momentum closest to the condition $\omega_P = \omega_Q/2$ is a surface normal to the direction $(0,0,q) , ~ q\approx 0.05 \AA^{-1}$. 

\section{Results}
\label{sec:res}
In addition to exploring the possibility of parametric down-conversion in solids across point groups, in Sec.~\ref{sec:mat_search}, here we highlight important physical signatures of the parametric coupling, and its dependence on the properties of the systems in question. For brevity, we will present results for representative candidates of three separate classes; 
the full list of the identified materials and their
symmetries can be found in \footnote{\href{https://github.com/danielkaplan137/spatioMaterials}{https://github.com/danielkaplan137/spatioMaterials} \label{github}}.


\subsection{Elemental solids}
\label{sec:elemental}
The simplest systems that host parametric coupling are the trigonal elemental solids, Se and Te. Both appear in the 32 ($D_3$) point group and their phonon spectra have been intensely studied due to their chiral phonons and the properties of their lattice dynamics in the presence of non-symmorphic symmetries \cite{teuchert1974symmetry,Pine1971,wendel1979lattice,Zhang2022}. From the symmetry constraints, we find that $E\times E^2$ is the only allowed parametric coupling. 
\begin{figure}[!ht]
    \centering
    \includegraphics[width=0.95\columnwidth]{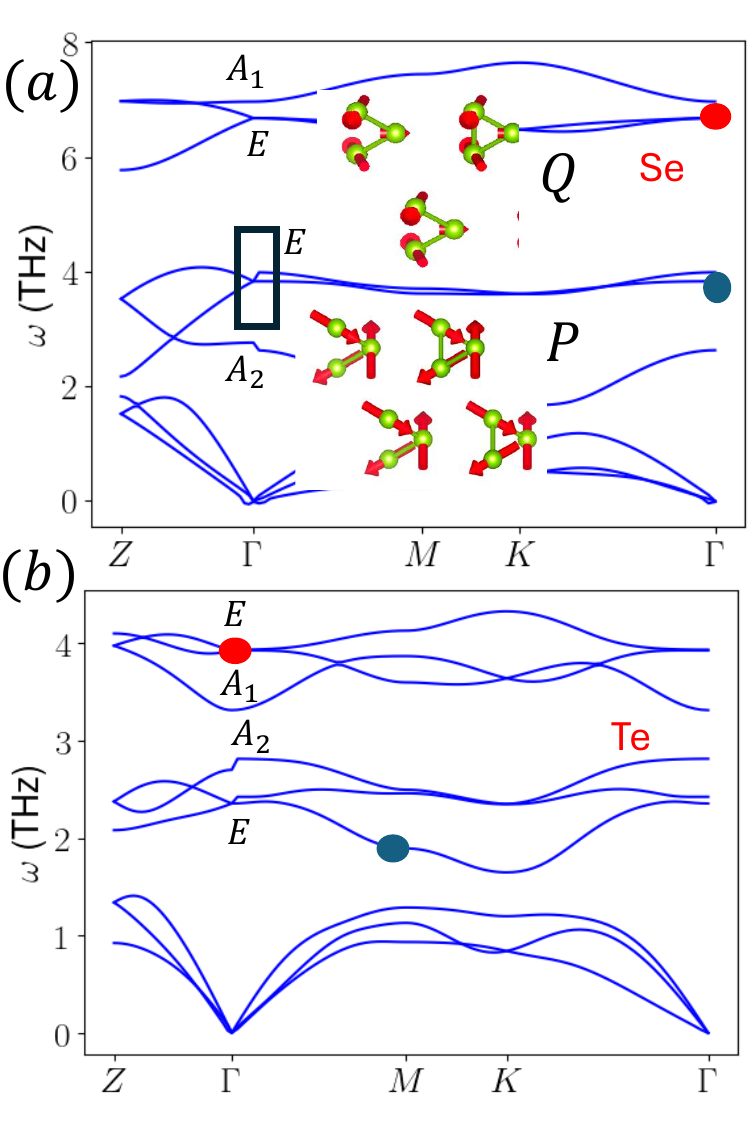}
    \caption{Phonon dispersion, representation inversion and LO-TO splitting in elemental Se and Te. (a) Results are presented for Se along $\Gamma -Z$ and in-plane direction, showing clear LO-TO splitting (inset, black box). The mode composition and symmetry order is unfavorable for parametric resonance. However, for Te, as shown in (b), the highest frequency mode is $E$ which may couple to a lower frequency $E$ mode, as shown via the red and blue dots, respectively.}
    \label{fig:fig4}
\end{figure}
The trigonal group allows us to explore two naturally occurring candidates: Se, Te, examining the effect of the heavier mass of Te and more delocalized valence manifold of $p$ electrons in Te versus Se. The two dispersions are shown in Fig.~\ref{fig:fig4}(a)-(b). We find that the change in the maximum phonon frequency consistent with the rough mass ratio enhancement of $\textrm{Te}/\textrm{Se} \approx 1.6$. However, the difference in the $p$-orbital structure of Te relative to Se leads to a change in the representation order: the largest phonon mode at $\Gamma$ is $E$ for Te but $A_1$ for Se, consistent with experimental observation \cite{powell1975lattice,Martin1976}. 

The chiral nature of the phonons in Se/Te is observed, and we plot the $Q$ and $P$ modes that emerge respectively. The atomic weight of the modes is qualitatively the same for both materials. Even though neither Se nor Te are polar, we observe LO-TO splitting as indicated in the inset to Fig.~\ref{fig:fig4}(a). 

For Se, we find that the resonance condition between $Q$, $P$ is not as comfortably satisfied as in Te. In particular, we note that the consequence of the LO-TO is the splitting of the doubly-degenerate $E$ mode in the in-plane direction. 
\begin{figure}[!ht]
    \centering
    \includegraphics[width=0.99\columnwidth]{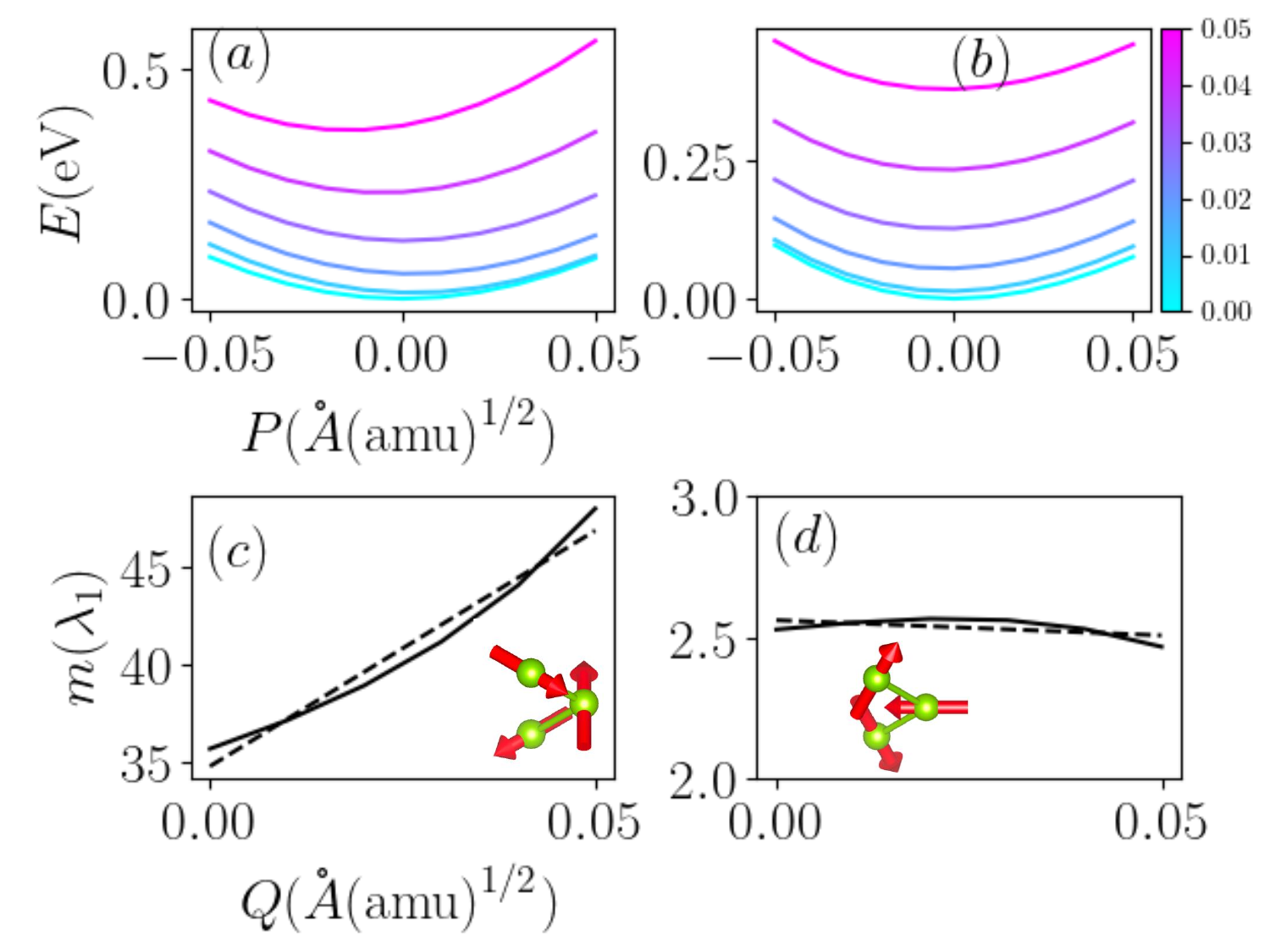}
    \caption{Coupling constants for Te between $E$ modes. (a)-(b) 
    Energy profiles of Te under $\mathbf{Q}$ excitations, which are color-coded according the scale on the right of(b); (a) shows the energy change for $\mathbf{Q}, \mathbf{P}$ modes of like chirality (clockwise), while (b) shows the energy change for dissimilar chirality. (c)-(d) Change in the mode mass, which determines the coupling $\chi^{1,2}$ as a function of $\mathbf{Q}$ mode amplitude. (c) Denotes the same chirality while (d) shows a nearly vanishing response for orthogonal chirality. }
    \label{fig:fig4p}
\end{figure}

As the chemical bonds in Se/Te are not ionic, the effective Born charge $Z$ is significantly reduced. However, the non-orthogonal nature of the effective charge matrix $Z_i^\alpha$ suggests that for circularly polarized light, light-induced chiral mode coupling can be realized despite the superficially small averaged $Z$. For Te, the effective charge matrix reads,
\begin{align}
    Z_1 = \left(
    \begin{matrix}
                 2.484 & 0.0 & 0.0 \\ 
         0.0 &   -2.497 &      2.097 \\
          0.0 & 2.572 & 0.0
    \end{matrix}
    \right), \\
    Z_2 = \left(
    \begin{matrix}
     -1.248	& -2.162 & -1.81674 \\
      -2.162   &   1.248 & -1.04890 \\
      -2.2	&  -1.286  &    0.0 
    \end{matrix}
    \right),
\end{align}
where the subscript denotes the one of the three equivalent $3a$ Te in the unit cell and $Z_3 = -(Z_1+Z_2)$. For Se, we find the components of $Z$ to be a factor of 4 smaller, making light-driven coupling less feasible. 

In order to meaningfully compute the threshold for parametric instability, the effective $Z$ for each atom is determined by evaluating the average value $Q^T_i Z_i Q$, which constitutes the projection of the phonon mode onto the atomic charge $Z_i$. To couple to the $E$ mode, we assume circularly polarized light along $y-z$. We take $\beta=0.1\omega_Q$ as above. 
\begin{table}[!ht]
\begin{tabular}{|p{3cm}|p{1cm}|p{1cm}| p{1cm}| p{1cm}| }
 \hline
 Element & $\chi^{1,2}$ & $\omega_Q$ & $q_0$ & $E_c$ \\
 \hline
 Se (iden. chirality) & 53.71 & 6.4 & 0.08 $\hat{z}$ & 14.5 \\ \hline
  Se (diff. chirality) & 0.003 & 6.4 & 
  0.601 $\hat{z}$ & $>$ 100 \\ \hline
  Te (iden. chirality) & 104.1 & 4.2 & 0.706 $\hat{y}$ & 1.79\\ \hline
  Te (diff. chirality) & 0.004 & 4.2 & 0.501 $\hat{z}$ & $>$ 100 \\ \hline
 \end{tabular}
 \caption{Summary of results for elemental chiral solids. The coupling constant is presented in units of $\textrm{meV} \AA^{-3}$, $\omega_Q$ is in $\textrm{THz}$, $q_0$ in $\AA^{-1}$ and $E_c$ in MVcm$^{-1}$. ``Iden." and ``diff." stand for indentical and different chiralities.}
 \label{tab:tab2}
 \end{table}

We find that the optimal ordering vector induced in Te lies closely to the $M$ point of the unfolded Brillouin zone. This suggests that the parametric coupling, with relevant constants presented
in Fig. ~\ref{fig:fig4p} and in Table ~\ref{tab:tab2}, may also drive commensurate transitions.

\subsection{Interatomic coupling}
\label{sec:interatomic}
\begin{figure}
    \centering
  \includegraphics[width=1.0\linewidth]{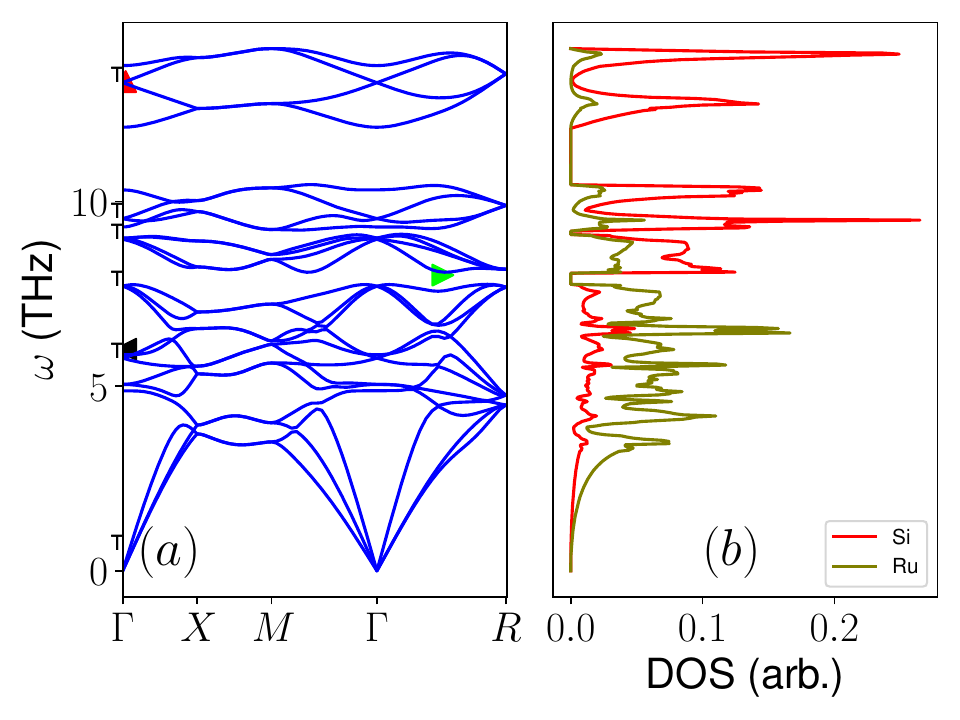}
    \caption{Phonon dispersion with irreps and DOS of RuSi. (a) Phonon energies are plotted along high-symmetry lines. The relevant $T$ modes are denoted with the letter $T$ next to their frequency at $\Gamma$. (b) Phonon density of states showing marked separation by Si and Ru branches. With the Red triangle we refer to the driven $T$ mode $\mathbf{Q}$. In Green, we assess the coupling to at $T$ primarily formed by Si nuclei; in Black, similarly, but for Ru atoms.}
    \label{fig:fig6}
\end{figure}

For solids with more than one atomic specie in the primitive cell, the phonon spectrum reveals mixing of vibrational modes between atoms of different orbital, chemical and local environments. We exploit this to investigate how the parametric coupling is affected between modes of intra-atomic versus inter-atomic vibrations.
We search for materials with sufficient separation in their atomic mass such that the atomic species may be separated from each other. The mode content must be large enough (i.e., the multiplicity of atomic positions) such that parametric instabilities may set on either for the intra- or inter- atomic modes.
Inspecting the database with Tab.~\ref{tab:1}, we find an optimal candidate in SiRu. This system belongs to the B20 (space group 194) family of cubic chiral crystals with the point group $23$ \cite{Mattheis_FeSi_1993, perring1999ruthenium,robredo2024multifold} which are now to host exotic electronic states. RuSi and OsSi (of the silicides) are semiconducting. Both Ru and Si occupy the $4a$ Wyckoff positions of this space group, and we represent them by the vector $\mathbf{R} = (\textrm{Si}_1,\textrm{Si}_2 \ldots,\textrm{Ru}_1,\ldots)^{T}$.
In Fig.~\ref{fig:fig6}(a) we plot the phonon dispersion of this compound and the associated, atom-resolved DOS in Fig.~\ref{fig:fig6}(b).

We find, as needed, two largely separate branches of Si and Ru vibrations. The double-degenerate $T$ mode at $\omega_Q = 13.215 \textrm{THz}$ (red triangle, Fig.~\ref{fig:fig6}(a)) is comprised mainly of Si motion, and represented by the vector, $u_1 = \left(-1,1,0,0,\ldots\right)$ and $u_2 = \left(0,0,1,-1,\ldots\right)$. Within the manifold of Si modes, we find a $T$ mode (green triangle, Fig.~\ref{fig:fig6}(a)) with predominately Si motion, $u = \left(-0.394,-0.394,-0.394, 0.394,\ldots\right)$. Similarly, and within rangeo of an instability is a predominantly Ru mode (black triangle, Fig.~\ref{fig:fig6}(a)), $u = (\ldots,0.284,-0.284,0.284,0.284)$. 

We find the coupling between $\btriangle$ ($\omega=13.2~\textrm{THz}$) and \textcolor{green}{$\btriangle$} ($\omega=8.7~\textrm{THz}$) to be nearly an order of magnitude stronger, than \textcolor{red}{$\btriangle$} and \textcolor{black}{$\btriangle$} ($\omega=5.8~\textrm{THz}$), with $\chi_{\color{red}{\btriangle}\color{green}{\btriangle}} = 12.71 \textrm{meV} \AA^{-3}= 8.8 \chi_{\color{red}{\btriangle}\color{black}{\btriangle}} $. This strongly supports a hypothesis presented in Ref.~\cite{Kaplan2025} that modes of similar atomic origin -- and in ferroelectrics, if they are related to the onset of the polar instability -- produce the strongest coupling. 

\subsection{Phases and Bulk vs Monolayer } 
\label{sec:bulk}

Recent advances in the engineering of ferroelectricity in low dimensions  \cite{li2023emergence,ramasamy2016solution,kamal_bandgaps_2016,wu2016intrinsic,xin2016few,jindal2023coupled,lipatov2022direct} suggests the investigation of the parametric coupling between phases and stacks of few layer to bulk van der Waals (vdW) materials. Most vdW exhibit polymorphs which are either inversion symmetric or polar, with the polarization strongly controlled by the stacking order \cite{li2024sliding, ouyang2025electrically}. Light-induced instability are only activated for those polytypes for which inversion symmetry is broken. As an example, we consider GaSe~\cite{kuhn1975crystal,Adler_GaSe_firstprinciples,ider2015thermochemistry} which has recently been actively studied for its emergent ferroelectricity \cite{li2023emergence}. We consider the four phases of GaSe which are nearly degenerate in energy \cite{Srour2018}. 

\begin{figure}
    \centering
  \includegraphics[width=1.0\linewidth]{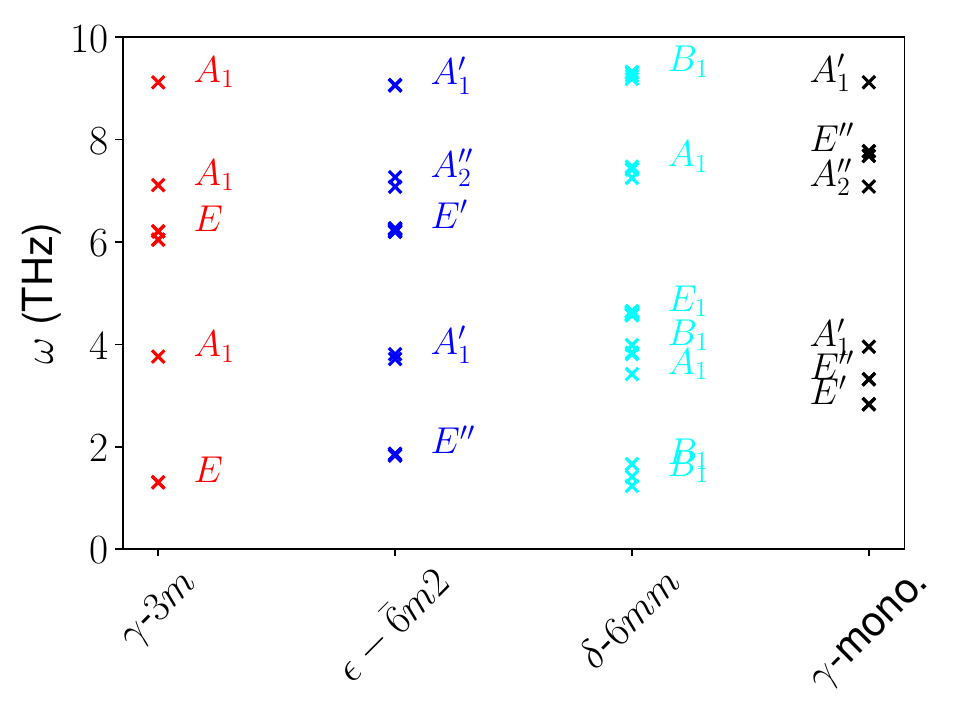}
    \caption{Phonon modes at $\Gamma$ and their symmetries for each of the phases of GaSe (excluding $\beta$, which is inversion symmetric). While most modes are roughly equivalent, there is a sharp transition in the mode order between $\gamma$-bulk and $\gamma$-monolayer which prevents parametric coupling in the monolayer. In addition, the $\delta$ phase contains imaginary frequencies (not shown) at this level of treatment in DFT, even though this phase has been observed \cite{kuhn1975crystal}. }
    \label{fig:fig7}
\end{figure}

In Fig.~\ref{fig:fig7} we show the phonon modes of different phases of GaSe. The broad similarity between the $\gamma$ and $\delta$ phases is also evinced in the symmetry irreps of the phonons.  However, the parametric coupling is more easily realized in the $\gamma$ phase: per Tab.~\ref{tab:1}, the TO $A_1$ mode at $7.11~\textrm{THz}$ may then downconvert to a mode of any symmetry. In the $\delta$ phase, however, only one light-driven phonon mode can parametrically de-excite, the $E'$ mode at $\textrm{6.05} ~\textrm{THz}$. This again illustrates that the coupling is sensitive to the point group symmetry and the observation of the parametric down-conversion may be used to constrain the local symmetry properties of the sample under illumination. 
The effect of thickness is quite considerable; since GaSe has been synthesized down to the monolayer, we inspected the phonon modes of a single layer. These differ dramatically from the $\gamma$ bulk. Due to the mode arrangement, we find that the $\gamma$-monolayer does not allow coherent de-excitation. This highlights the thickness sensitivity of the parametric coupling, in approaching the 2D limit. 

Lastly, we observed that the wurtzite analogue of GaSe ($\delta$) to have the strongest nonlinear coupling. However, we caution that this may be an overestimation, as within DFT (see App.~\ref{App:A}) without vdW dispersive corrections, imaginary frequencies were found in the structure, suggesting it is unstable at the PBE level (although observed experimentally, at finite temperature). A detailed calculation using more sophisticated methods will be carried out in future work.  We note that the incommensurate ordering wavevector is experimentally
accessible by time-resolved diffraction \cite{Kaplan2025}; in the case
of the two-dimensional candidates ultrafast electron diffraction \cite{Filipetto_timeresolved_2022} is a  possible probe.

A summary of the properties of these phases and their nonlinear coupling is presented in Tab.~\ref{tab:tab3}. 

\begin{table}[!ht]
\begin{tabular}{|c |c|c| c|}
 \hline
 GaSe phase (point group)  & $\chi^{1,2}$ & $\omega_Q$ & $q_0$ \\
 \hline
 $\beta$ (6/mmm) & 0 & &   \\ \hline
  $\epsilon$ ($\bar{6}$m2) & 1.2 & 6.05 &   0.05 $\hat{x}$  \\ \hline
  $\gamma$ (R3m) & 4.45 & 7.11 & 0.12  $\hat{x}$ \\ \hline
$\delta$ (6mm) & 17.5 & 7.08 & 0.04  $\hat{x}$ \\ \hline
  $\gamma$-mono & - &  &   \\ \hline
 \end{tabular}
 \caption{Summary of results for elemental chiral solids. The coupling constant is presented in units of $\textrm{meV} \AA^{-3}$, $\omega_Q$ is in $\textrm{THz}$, $q_0$ in $\AA^{-1}$.}
 \label{tab:tab3}
 \end{table}
\section{Discussion}
\label{sec:disc}
In this work, we have carried out a systematic \textit{ab-initio} study of materials towards the realization of light-driven parametric instabilities in solids. We have decomposed the nonlinear phonon interaction through symmetries, and have 
deduced the point groups for which nonlinear interactions, which allow light-driven parametric instabilities,
are symmetry-allowed. Next we have surveyed the full materials landscape revealing an abundance of potential candidates. At present, \matpro contains 2078 materials which would in principle permit these nonlinear interaction to arise. However, as
 we outlined in Secs.~\ref{sec:theory}-\ref{sec:mat_search},
the parametric instability only occurs only under specific conditions. As a result, we initially narrowed the search to systems with precomputed phonon dispersion; this established 278 candidates. After this, we described in full the computational procedure for obtaining the coupling with any DFT suite capable of computing phonon dispersion. 

Our results indicate that traditional ferroelectric insulators with ionic chemical bonding are prime candidates for light-driven instabilities; their large effective Born charges ($Z$) strongly
impact the threshold field for the instability. This closely follows the principles outlined in Ref.~\cite{Kaplan2025} and explored in Ref.~\cite{Subedi2015} for ultrafast switching of ferroelectric states in polar materials. 
However, the strength of the nonlinear coupling is revealed to be strongly dependent on other factors. In Sec.~\ref{sec:interatomic} we demonstrate that the coupling significantly benefits from phonons interacting within the same atomic sector -- i.e., originating from the atomic modes of the distortion that created the ferroelectric state from a symmetry parent. We also present this property in RuSi, a cubic chiral crystal of the FeSi family \cite{Mattheis_FeSi_1993,kloc1995preparation}. This suggests that certain topological properties found in cubic silicides \cite{schroter2019chiral,robredo2024multifold} could be successfully tuned and manipulated with light-induced spatiotemporal
order. Finally, we show how the coupling is strongly sensitive to the phase polymorphism of many vdW stacked materials. Focusing on the prototypical GaSe ferroelectric, we demonstrate that the coupling is sensitive to the atomic content of different phases and the symmetry of the phonon modes, which differs between phases. This would promote the parametric instability as a probe of structure and symmetry in solids, similar to way in which nonlinearities have been used in other contexts for the tomography of electronic states \cite{suarez2024non,kaplan2025quantum}.

We briefly comment here on certain approximations made in discussing the theory of the spatiotemporal order: we specifically neglected couplings of the form $Q^2P^2$ which are in principle allowed for any symmetry. From a perturbative sense, these are of course smaller than the leading $P^2Q$ if the latter is allowed by symmetry. 
However, such a term contributes to the softening of the mass of either $P$ or $Q$ which can lead to instabilities in systems with soft phonons \cite{zhuang2023light}.
As the light-induced order corresponds to an incommensurate translational symmetry-breaking that oscillates with time, it is ideally detected through time-resolved diffractive probes, such as ultrafast electron diffraction \cite{Filipetto_timeresolved_2022}. 
This was recently employed, for example, in the visualization of charge density wave melting due to light \cite{kogar2020light,lee2025topologicalphasetransitionhidden}.

Looking ahead, our proposed methodology as summarized in Fig.~\ref{fig:intro} is directly applicable to high-throughput searches targeting optimization of $\chi^{1,2}$ for dynamical symmetry-breaking and structural engineering. Combined with topological properties \cite{Bordoloi2025} and softened barriers to phase transitions, such as in sliding ferroelectric bilayers \cite{Wei2025}, this approach may enable dynamical symmetry-breaking as a switchable platform for memory and computational applications. In metals with broken inversion symmetry, we expect substantial influence on electronic properties through 
electron-phonon coupling; combined with correlations \cite{Lau2025,coulter2025electron}, we expect the spatiotemporal order to serve as a new method of controlling electronic properties as well. 

\begin{acknowledgements}
We are grateful to A. Cavalleri and I.I. Mazin for stimulating discussions. 
DK acknowledges support from the Abrahams Postdoctoral Fellowship of the center for Materials Theory, Rutgers University and the Zuckerman STEM Fellowship. DK and PC thank the Center for Computational Quantum Physics at the
Flatiron Institute for hospitality during the course of this project. The Flatiron Institute is a division of the Simons Foundation.
\end{acknowledgements}

\appendix
\section{Computational details}
\label{App:A}
All calculations in this work were carried out using Quantum Espresso (QE) \cite{QE-2009,QE-2017} and phonon calculations were performed using the methods outlined in Ref.~\cite{Baroni2001}, specifically density functional perturbation theory (DFPT). Extensive use was made of AiiDA \cite{huber2020aiida,uhrin2021workflows} for the preparation of QE input files. All $k$-space grids were chosen using the ``balanced" protocol, such that k-point spacing along a lattice vector was at or below $0.15 \AA^{-1}$, with the sampling determined by the Monkhorst-Pack method \cite{MP_grid} . The xc functional used was PBEsol \cite{PBEsol}. Pseudopotentials were obtained for each task using the SSSP PBEsol efficiency v1.3.0 \cite{prandini2018precision}. Cutoffs for the charge density and wavefunction plane wave expansions were taken directly from SSSP. Most pseudopotentials were of the PAW type, and originated from \textrm{pslibrary} \cite{dal2014pseudopotentials}. Spin orbit coupling was not considered. Structures obtained from \matpro were relaxed until forces were $ < 10^{-4}~\textrm{Ry}/\textrm{atom}$. $k$-space paths were sampled along suggested high-symmetry directions using the approach proposed in Ref.~\cite{Hinuma_2017}. In order to compute the simple products of point group irreps., we used the database provided by the Bilbao Crystallographic Server \cite{aroyo2006bilbao}.

All materials treated here are insulators, and thus calculations were done at fixed occupations without smearing. 
\section{Post-processing}
The full list of materials screened in the present work is available at the repository \cite{Note1}. There, we have placed post-processing scripts which directly extract information and prepare input files for QE. 
\bibliography{main.bbl}

\end{document}